\documentclass[a4paper,12pt]{article}
\usepackage{amsmath}
\usepackage{epsfig,subfigure}
\usepackage{cite}
\date{}
\begin{document}
\title{
\large{
\textbf{Influence of the flow on the anchoring of nematic
liquid crystals on a Langmuir-Blodgett monolayer 
studied by optical second-harmonic generation
}
}
}\normalsize
\author{\large
Valentina S. U. Fazio$^{1,2}$\footnote{e-mail: fazio@fy.chalmers.se}, 
Lachezar Komitov$^{1}$\\ 
\large Christian Rad\"uge$^{2}$, Sven T. Lagerwall$^{1}$,\\
\large Hubert Motschmann$^{2}$\\
$^{1}$\textit{\small{
Department of Microelectronics and Nanoscience,
Liquid Crystal Physics,
}}\\
\textit{\small{Chalmers University of Technology \&
G\"oteborg University,}}\\ 
\textit{\small{SE-41296 G\"oteborg, Sweden
}} \\
$^{2}$\textit{\small{
Max-Plank-Institute of Colloids and Interfaces}}\\
\textit{\small{
D-14476 Golm/Potsdam, Germany
}}
}
\normalsize
\maketitle
\begin{abstract}
The influence of capillary flow on the alignment of the nematic 
liquid crystal MBBA on fatty acid Langmuir-Blodgett monolayers was 
studied by optical second-harmonic generation. 
The surface dipole sensitivity of the technique allows probing the
orientation of the first liquid crystal monolayer in the presence of 
the liquid crystal bulk.
It was found that capillary flow causes the first monolayer of liquid 
crystal molecules in contact with the fatty acid monolayer to be 
oriented in the flow direction with a large pretilt (78 degrees), 
resulting in a quasi-planar alignment with splay-bend deformation of 
the nematic director in the bulk.
The large pretilt angle also suggests that the Langmuir-Blodgett film 
itself is affected by the flow.
The quasi-planar flow-induced alignment was found to be metastable. 
Once the flow ceases, circular domains of homeotropic orientation nucleate in 
the sample and expand until the whole sample becomes homeotropic.
This relaxation process from flow-induced quasi-planar to 
surface-induced homeotropic alignment was also monitored by SHG.
It was found that in the homeotropic state the first nematic layer 
presents a pretilt of 38 degrees almost isotropically 
distributed in the plane of the cell, with a slight preference for the
direction of the previous flow. 
\end{abstract}
\section{Introduction}
Substrate-induced alignment of liquid crystals (LCs) is of crucial 
significance for LC display devices.
Liquid crystal molecular ordering at an interface can be 
significantly different from that in the bulk\cite{GuyHsiShe86, 
SkaAleBarZum98}.
This problem is not only of fundamental interest, but also of 
practical importance.
In the past decade numerous studies have been dedicated to the 
phenomenon of anchoring of liquid crystals on solid substrates
\cite{Jerome91}, but the mechanisms by which a surface imposes a certain 
orientation to a liquid crystal is far from being well understood.
Most of the research in this field has been done with nematic 
liquid crystals (NLCs) which are the simplest liquid crystalline phase
formed by elongated molecules that, in average, are oriented parallel 
to each other\cite{deGennes}.
Depending on the treatment, a solid surface may orient the LC in a 
certain direction, called anchoring direction, which is not 
necessarily parallel to that of the first few molecular layers in 
contact with the surface\cite{ZhuMarShe94}.

In the bulk of a nematic liquid crystal the molecules are aligned 
along a preferred direction called director, $\mathbf{n}$,
in a nonpolar manner\cite{deGennes}, i.e.
\begin{equation}
\mathbf{n} \leftrightarrow -\mathbf{n}.
\label{10}
\end{equation}
This leads to a macroscopic structure which is centrosymmetric, where 
second-harmonic generation (SHG) is forbidden (in the dipole approximation).
The second-harmonic (SH) nonlinear polarization is proportional to the 
SH nonlinear susceptibility $\chi^{(2)}$ which, under the dipole approximation, 
vanishes for a medium with inversion symmetry\cite{PrasadWilliams}.
At an interface, however, the inversion symmetry is broken and 
$\chi^{(2)} \neq 0$ \cite{Shen}.
This makes SHG a surface-sensitive technique which can be used for 
instance to determine the orientational distribution of a surface 
monolayer of LC molecules in absence or in presence of the LC bulk.

\begin{figure}
\begin{center}
\epsfig{file=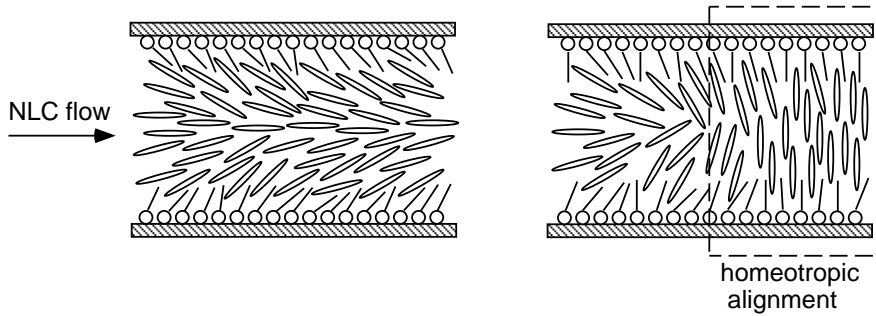, width=0.85\textwidth}
\caption{During filling (left) the NLC adopts a  
quasi-planar alignment with splay-bend deformation and preferred 
orientation along the filling direction (left).
Once the flow stops, domains of homeotropic orientation nucleate 
and expand in the sample (right). 
\label{cellfilling_relax}
}
\end{center}
\end{figure}

Generally, capillary flow influences the orientation induced by a solid 
substrate on a nematic liquid crystal\cite{BarChuKre92}, in particular 
when the anchoring in weak, as in the case of homeotropic 
alignment by surfactants.
In the case of aligning Langmuir-Blodgett (LB) monolayers 
of fatty acids it is found
\cite{FazKomLag98a} that during capillary flow in the nematic phase 
the liquid crystal (MBBA) is quasi-planarly oriented along the flow 
direction with a bulk splay-bend deformation (see Figure 
\ref{cellfilling_relax}).   
As soon as the flow ceases, circular homeotropic domains  
(orientation along the substrate normal) nucleate in the sample and 
expand until the whole sample becomes homeotropic.
The speed of expansion of the homeotropic domains is found to 
be very sensitive to the fatty acid monolayer\cite{FazNanKomXX}: 
it decreases with increasing the length of the fatty acid alkyl 
chain and changes substantially for small variations in the chain, 
like the addition of two CH$_{3}$ groups.
This fact suggests that the LB film plays a role in the process of 
relaxation from the flow-induced quasi-planar to the surface-induced 
homeotropic alignment.
The achieved homeotropic alignment is generally very good\cite{FazKomLag98b}
but the surface preserves a memory of the flow which can be revealed 
after that electric-field--induced breaking of the homeotropic anchoring 
takes place in the sample\cite{FazKom99}.

In this work we used optical SHG to study the alignment of a nematic 
liquid crystal (5CB) on a LB monolayer of stearic acid (C18) during and 
after capillary flow.
The technique allows us to evaluate the orientational distribution of 
the nematic molecules of the first monolayer in contact with the aligning 
LB film\cite{note00} and thus to study indirectly the effect of the capillary 
flow on the LB monolayer itself.

\section{Theory}
In SHG the second-harmonic signal is generated by the nonlinear 
polarization
\begin{equation}
P^{(2)}(2 \omega) = \bar{\chi^{(2)}} : E(\omega) E(\omega),
\label{12}
\end{equation}
where $E(\omega)$ is the excitation field at the fundamental 
frequency $\omega$, and $\bar{\chi^{(2)}}$ is the second-order 
nonlinear susceptibility tensor.
In the case of surface SHG $P^{(2)}(2 \omega)$ and $\bar{\chi^{(2)}}$
are the surface second-order nonlinear
polarization and susceptibility tensor, 
respectively.
The intensity $\mathcal{I}(2 \omega)$ of the generated second-harmonic 
light is given by\cite{Boyd}:
\begin{equation}
\mathcal{I}(2 \omega) \propto 
\left| \mathbf{e}_{2 \omega} \bar{\chi^{(2)}} : \mathbf{e}_{\omega}
\mathbf{e}_{\omega} \right|^{2} \, \mathcal{I}^{2}(\omega),
\label{13}
\end{equation}
where $\mathcal{I}(\omega)$ is the fundamental light intensity and 
$\mathbf{e}_{2 \omega}$ and $\mathbf{e}_{\omega}$ represent the 
polarization of the SH and of the fundamental light, respectively. 

If LC molecules are adsorbed on a surface, they generally form a 
polar layer as the surface imposes a preferred molecular alignment.
In this case, the nonlinear susceptibility of the substrate is 
negligible with respect to that of the LC monolayer. 
Moreover, if the LC possesses a dominant hyperpolarizability
$\beta^{(2)}_{\xi\xi\xi}$ along the long molecular axis 
$\mathbf{\xi}$,
the second-order nonlinear optical susceptibility coefficients can be 
written in the simple form\cite{FelCheShe91}:
\begin{equation}
d_{xyz} 
= \frac{1}{2} \, \chi^{(2)}_{xyz} 
= \frac{1}{2} \, N_{s} \, \left\langle
(\mathbf{x} \cdot \mathbf{\xi})
(\mathbf{y} \cdot \mathbf{\xi})
(\mathbf{z} \cdot \mathbf{\xi})
\right\rangle \, \beta^{(2)}_{\xi\xi\xi},
\label{20}
\end{equation}
where $N_{s}$ is the surface density of the molecules,
$\mathbf{x}$, $\mathbf{y}$ and $\mathbf{z}$ are the substrate 
coordinates as shown in Figure \ref{schemaSHGflow}, and 
$\left\langle \right\rangle$ denotes the average over the 
molecular orientational distribution function, $f_{\theta\phi}$, 
where $\theta$ and $\phi$ are the polar and azimuthal angles 
of $\mathbf{\xi}$ with respect to the substrate coordinate system
(see Figure \ref{schemaSHGflow}).

For an isotropic distribution of molecules in a monolayer there are 
only two independent non-vanishing components of $\bar{d}$\cite{FelCheShe91}, 
namely $d_{zzz}$ and $d_{iiz}$, where $i = x,y$:
\begin{eqnarray}
d_{zzz} &=& \frac{1}{2} \, N_{s} \, \left\langle \cos^{3}\theta 
\right\rangle \, \beta^{(2)}_{\xi\xi\xi},\\ \nonumber
d_{iiz} &=&  \frac{1}{4} \, N_{s} \, \left\langle \sin^{2}\theta 
\, \cos \theta \right\rangle \, \beta^{(2)}_{\xi\xi\xi},
\label{30}
\end{eqnarray}
where $\theta$ is the angle between $\mathbf{\xi}$ and $\mathbf{z}$ 
(polar angle in Figure \ref{schemaSHGflow}).
For molecules having a preferred alignment along $\mathbf{x}$ 
(as, for instance, a flow-induced orientation, with the flow in the 
$\mathbf{x}$ direction) the 
independent components of the $\bar{d}$-tensor are six\cite{FelCheShe91}:
\begin{eqnarray}
d_{zzz} &=& \frac{1}{2} \, N_{s} \, 
\left\langle \cos^{3}\theta \right\rangle \, \beta^{(2)}_{\xi\xi\xi},
 \nonumber \\
d_{xxx} &=& \frac{1}{2} \, N_{s} \, 
\left\langle \sin^{3}\theta \right\rangle \, 
\left\langle \cos^{3}\phi  \right\rangle \, \beta^{(2)}_{\xi\xi\xi},
 \nonumber \\
d_{zyy} = d_{yzy} = d_{yyz} &=& \frac{1}{2} \, N_{s} \, 
\left\langle \cos\theta - \cos^{3}\theta \right\rangle \, 
\left\langle 1 - \cos^{2}\phi  \right\rangle \, \beta^{(2)}_{\xi\xi\xi},
 \nonumber \\
d_{zxx} = d_{xzx} = d_{xxz} &=& \frac{1}{2} \, N_{s} \, 
\left\langle \cos\theta - \cos^{3}\theta \right\rangle \, 
\left\langle \cos^{2}\phi  \right\rangle \, \beta^{(2)}_{\xi\xi\xi},
 \nonumber \\
d_{xzz} = d_{zxz} = d_{zzx} &=& \frac{1}{2} \, N_{s} \, 
\left\langle \sin\theta - \sin^{3}\theta \right\rangle \, 
\left\langle \cos\phi  \right\rangle \, \beta^{(2)}_{\xi\xi\xi},
 \nonumber \\
d_{xyy} = d_{yxy} = d_{yyx} &=& \frac{1}{2} \, N_{s} \, 
\left\langle \sin^{3}\theta \right\rangle \, 
\left\langle \cos\phi - \cos^{3}\phi \right\rangle \, \beta^{(2)}_{\xi\xi\xi},
\label{40}
\end{eqnarray}
where $\phi$ is the 
molecular azimuthal angle, the angle that the projection of the long 
molecular axis makes with the $\mathbf{x}$ axis, 
as defined in Figure \ref{schemaSHGflow}.
In Equation \ref{40} the distribution functions for $\theta$ and $\phi$,
$f_{\theta}$ and $g_{\phi}$ are 
assumed to be independent\cite{FelCheShe91} (this is however not 
always the case\cite{note01}), i.e., $f_{\theta\phi} = 
f_{\theta} \cdot g_{\phi}$.

The distribution functions for the polar angle $\theta$ (polar 
distribution functions) generally used are $\delta$- and Gaussian 
functions:
\begin{eqnarray}
f_{\theta} &=& F \, \delta(\theta - \theta_{0}),  \label{50} \\
f_{\theta} &=& F \, e^{-\frac{(\theta - 
\theta_{0})^{2}}{2\,\sigma^{2}}},   \label{60}
\end{eqnarray}
where $\theta_{0}$ and $\sigma$ are the average molecular tilt and its 
variance.
Combining the $d$-coefficients in Equation \ref{40} leads to
\begin{eqnarray}
\frac{\left\langle \sin^{3}\theta \right\rangle_{f_{\theta}}}
{\left\langle \sin\theta \right\rangle_{f_{\theta}}} &=&
\frac{d_{xxx} + d_{xyy}}{d_{xxx} + d_{xyy} + d_{xzz}}, \label{70} \\
\frac{\left\langle \cos^{3}\theta \right\rangle_{f_{\theta}}}
{\left\langle \cos\theta \right\rangle_{f_{\theta}}} &=&
\frac{d_{zzz}}{d_{zzz} + d_{zxx} + d_{zyy}}, \label{80} 
\end{eqnarray}
where $\left\langle  \right\rangle_{f_{\theta}}$ is the average over 
the polar distribution function. 
Thus, by measuring the $d$-coefficients with a SHG experiment 
$\theta_{0}$ and $\sigma$ can be calculated. 

The azimuthal distribution function $g_{\phi}$ can be built up 
from a Fourier series\cite{FelCheShe91}:
\begin{equation}
g_{\phi} = \sum_{n} \, a_{n} \, \cos n\phi.
\label{90}
\end{equation}
From the remaining independent combinations of the $d$-coefficients 
in Equation \ref{40} the Fourier coefficients $a_{n}$ can be 
determined up to the third order.

Finally, from the azimuthal distribution function $g_{\phi}$ a surface 
in-plane order parameter, $S_{\text{s}}$, can be calculated.
Assuming independent distributions for $\theta$ and $\phi$, 
$S_{\text{s}}$ can be written as
\cite{KaiNakKan87,BarAerHolDam92}:
\begin{equation}
S_{\text{s}} = \left\langle \cos 2 \phi 
\right\rangle_{g_{\phi}} 
= \frac{1}{2} \, a_{2} = 
\frac{d_{zxx} - d_{zyy}}{d_{zxx} + d_{zyy}},
\label{100}
\end{equation}
where the average is done over the area illuminated by the laser beam.
In case of perfect order in the liquid crystal
monolayer $S_{\text{s}} = 1$,
while for a completely isotropic azimuthal distribution of molecules 
in the monolayer $S_{\text{s}} = 0$.

\section{Experiment}
\begin{figure}
\begin{center}
\epsfig{file=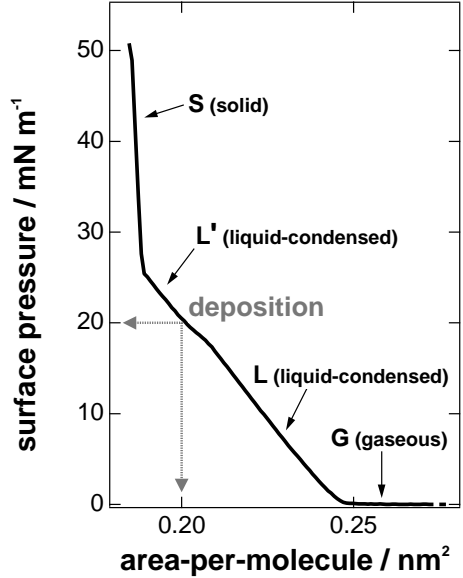, width=0.5\textwidth}
\caption{Isotherm of stearic acid (C18).
The monolayers were deposited at a surface pressure of 
20\,mN\,m$^{-1}$ and at an area-per-molecule of 0.2\,nm$^{2}$.
Transfer ratio of depsition was close to one.
\label{C18isot}
}
\end{center}
\end{figure}

LB monolayers of stearic acid (C18) were deposited onto ITO coated 
glass plates that were assembled in sandwich cells 10\,$\mu$m thick.
The isotherm of stearic acid is shown in Figure \ref{C18isot} 
where the deposition conditions are indicated.
The monolayers were deposited with a transfer ratio close to one.

\begin{figure}
\begin{center}
\epsfig{file=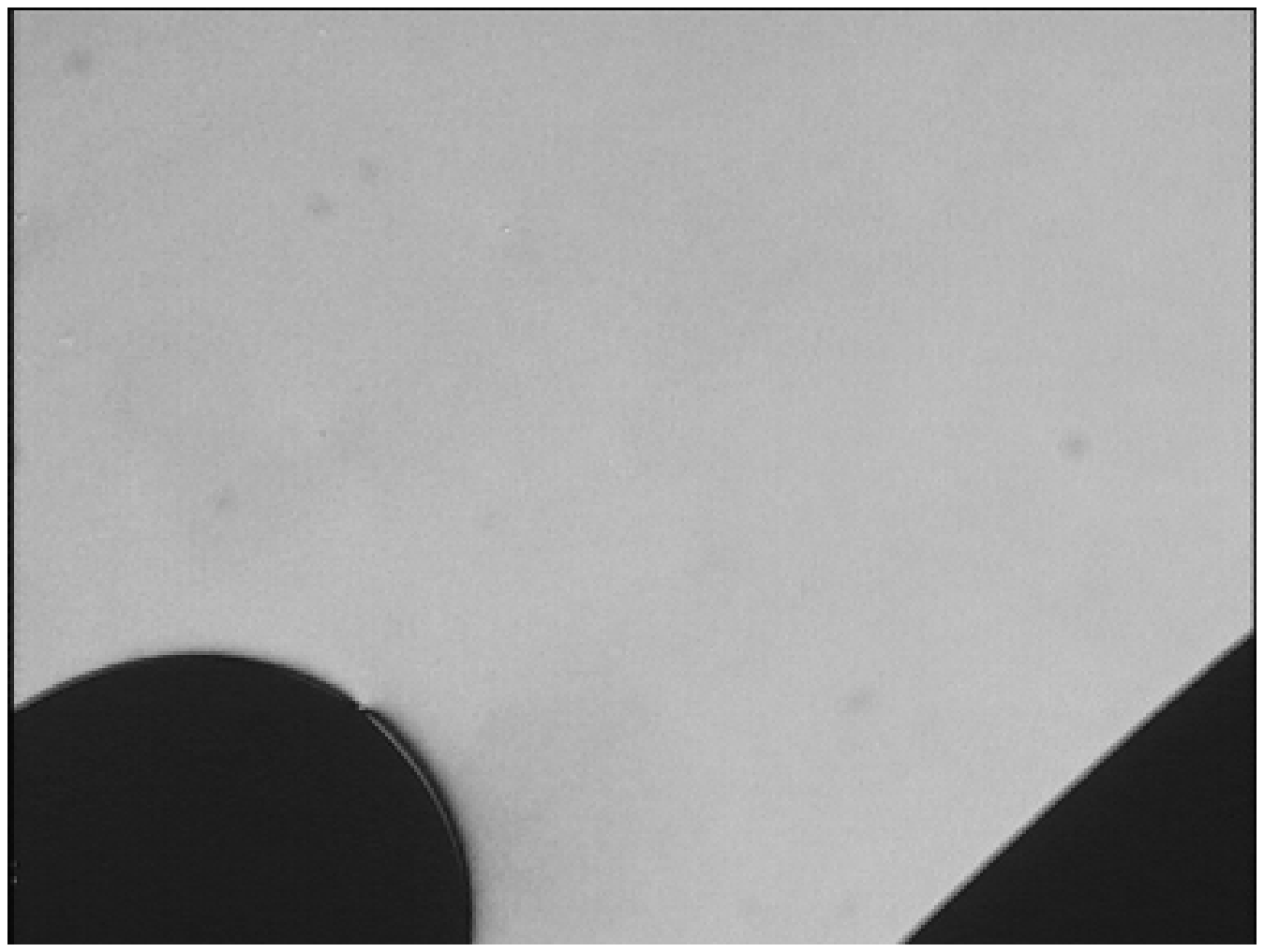, width=0.30\textwidth}
\hspace{0.03\textwidth}
\epsfig{file=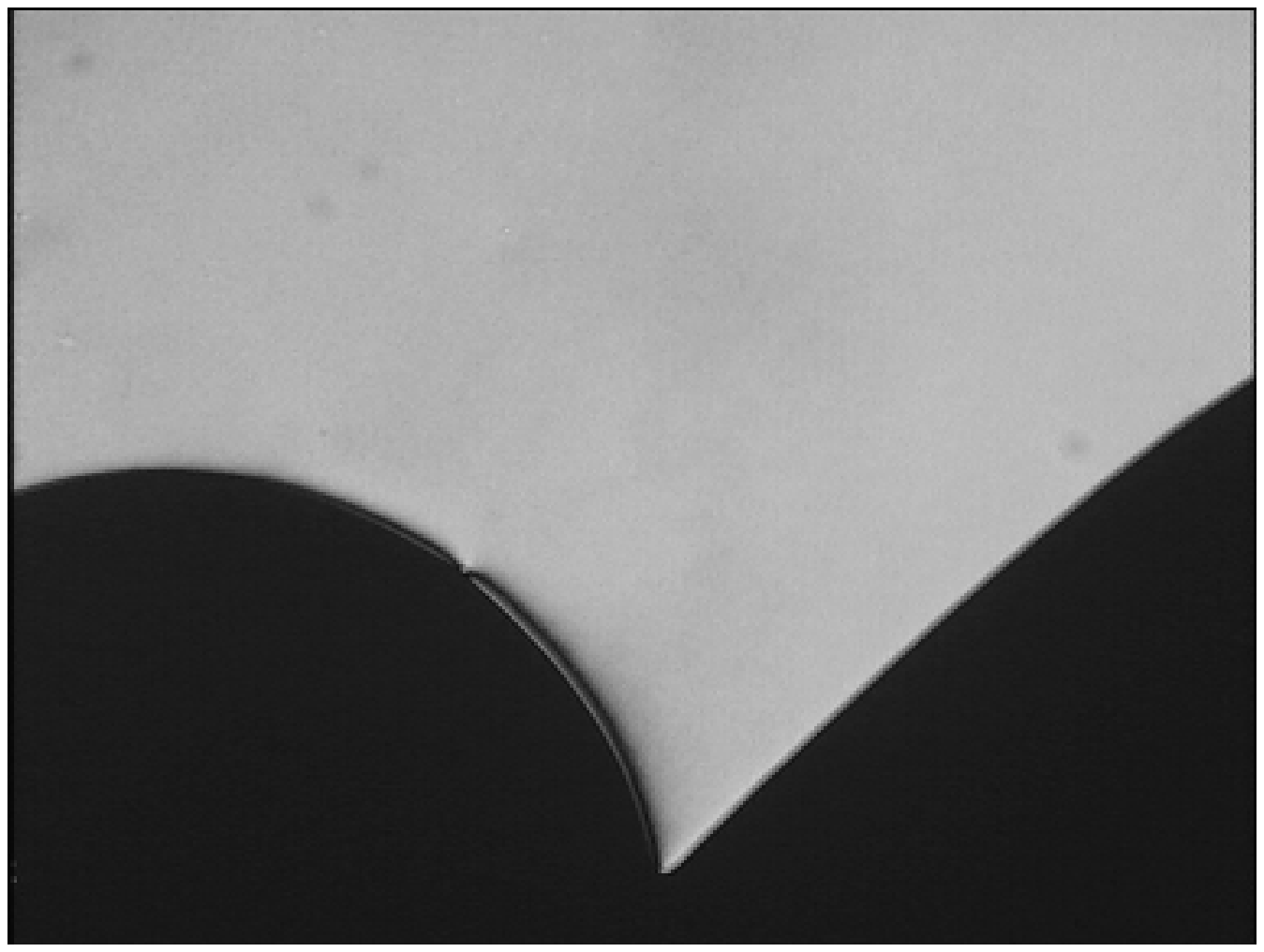, width=0.30\textwidth}
\hspace{0.03\textwidth}
\epsfig{file=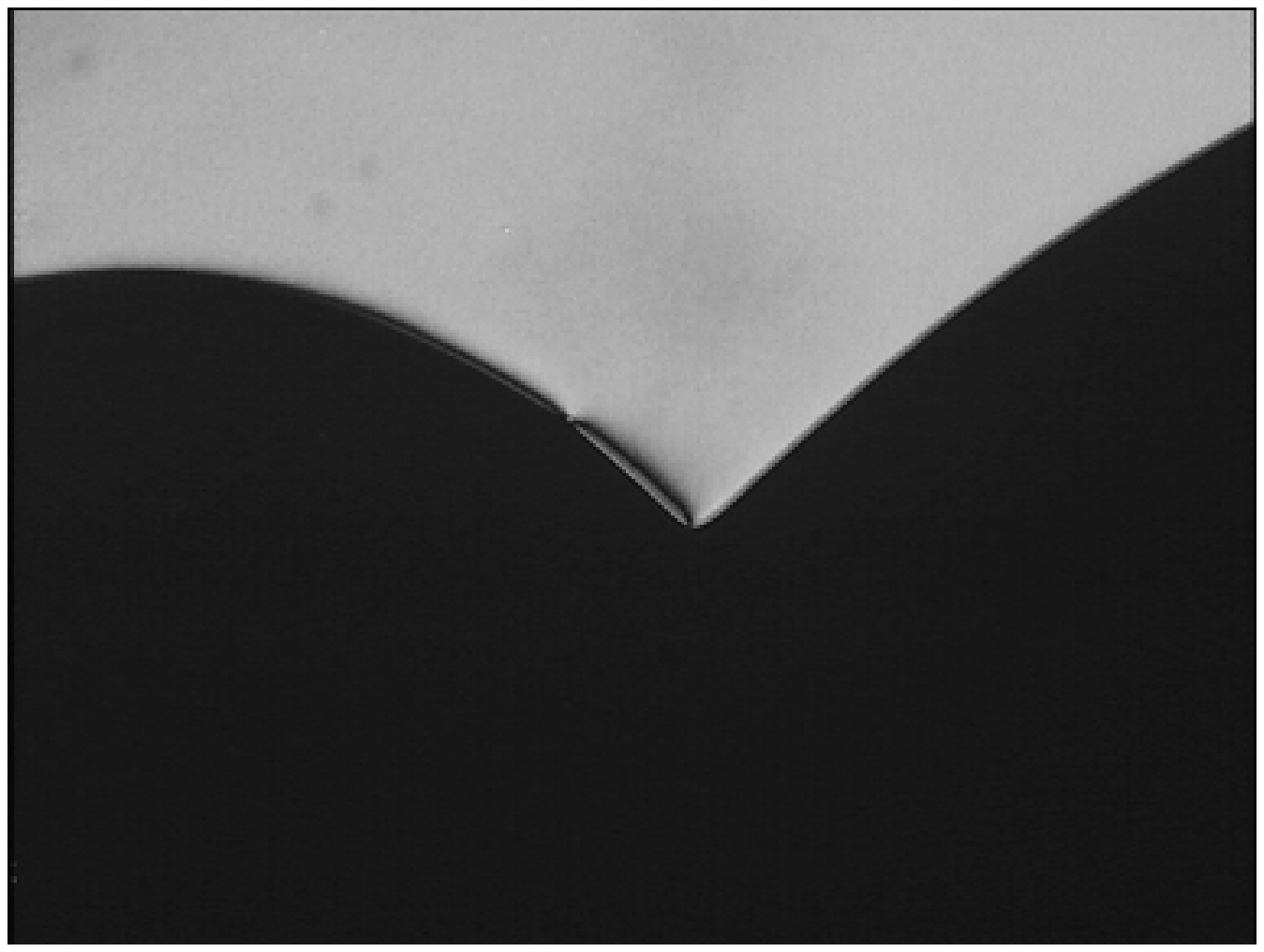, width=0.30\textwidth}
\caption{Cell between crossed polarizers. 
The 5CB flow has ceased and the homeotropic domain (black) expands in the 
quasi-planar one (gray).
The cell thickness is 10\,$\mu$m.
The three pictures are taken at a time interval of 20\,s.
Magnification 10.
\label{relax_ex}
}
\end{center}
\end{figure}

The cells were filled with 5CB (NC$\Phi \Phi$C$_{5}$H$_{11}$) in the nematic 
phase (5CB exhibits a nematic phase between 24$^{\circ}$C and 35$^{\circ}$C).
During capillary flow the NLC adopts a quasi-planar alignment with 
splay-bend deformation and preferred orientation along the flow 
direction (Figure \ref{cellfilling_relax}).
As in the case of MBBA\cite{FazKomLag98a}, the flow-induced 
alignment is metastable and transforms into homeotropic via nucleation 
and expansion of homeotropic domains.
An example of how the homeotropic domains expand in the quasi-planar 
one after the capillary flow has ceased is shown in Figure 
\ref{relax_ex}.
%
%\psdraft
\psfull
5CB also possesses a dominating hyperpolarizability along the long 
molecular axis $\mathbf{\xi}$, which makes it 
suitable for SHG studies.

\begin{figure}
\begin{center}
\epsfig{file=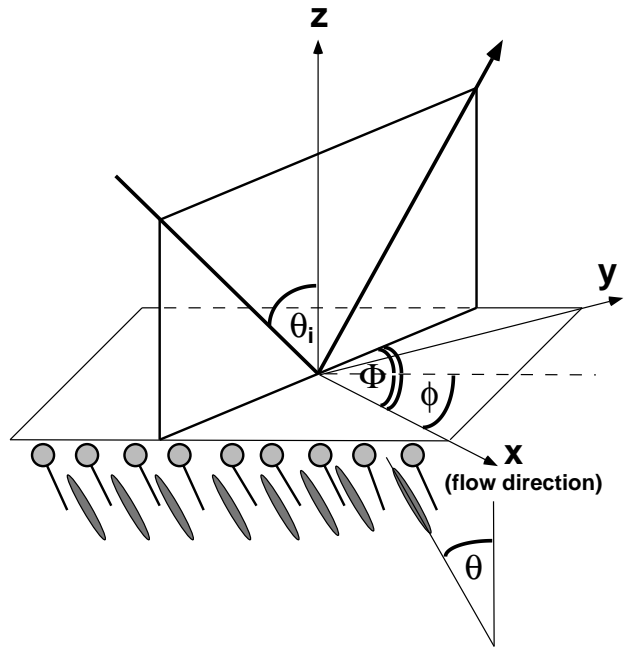, width=0.6\textwidth}
\caption{Schematic representation of the experimental geometry.
$\theta_{\text{i}}$ is the incidence angle (55 deg).
The flow is along the $\mathbf{x}$-direction.
$\theta$ and $\phi$ are the polar and azimuhal angles, respectively
of the LC monolayer molecules.
$\Phi$ is the angle between the plane of incidence and the direction 
of the flow.
The $xy$-plane is the substrate plane and the $\mathbf{z}$ direction is the 
\lq\lq homeotropic direction\rq\rq. 
\label{schemaSHGflow}
}
\end{center}
\end{figure}

\begin{figure}
\begin{center}
\epsfig{file=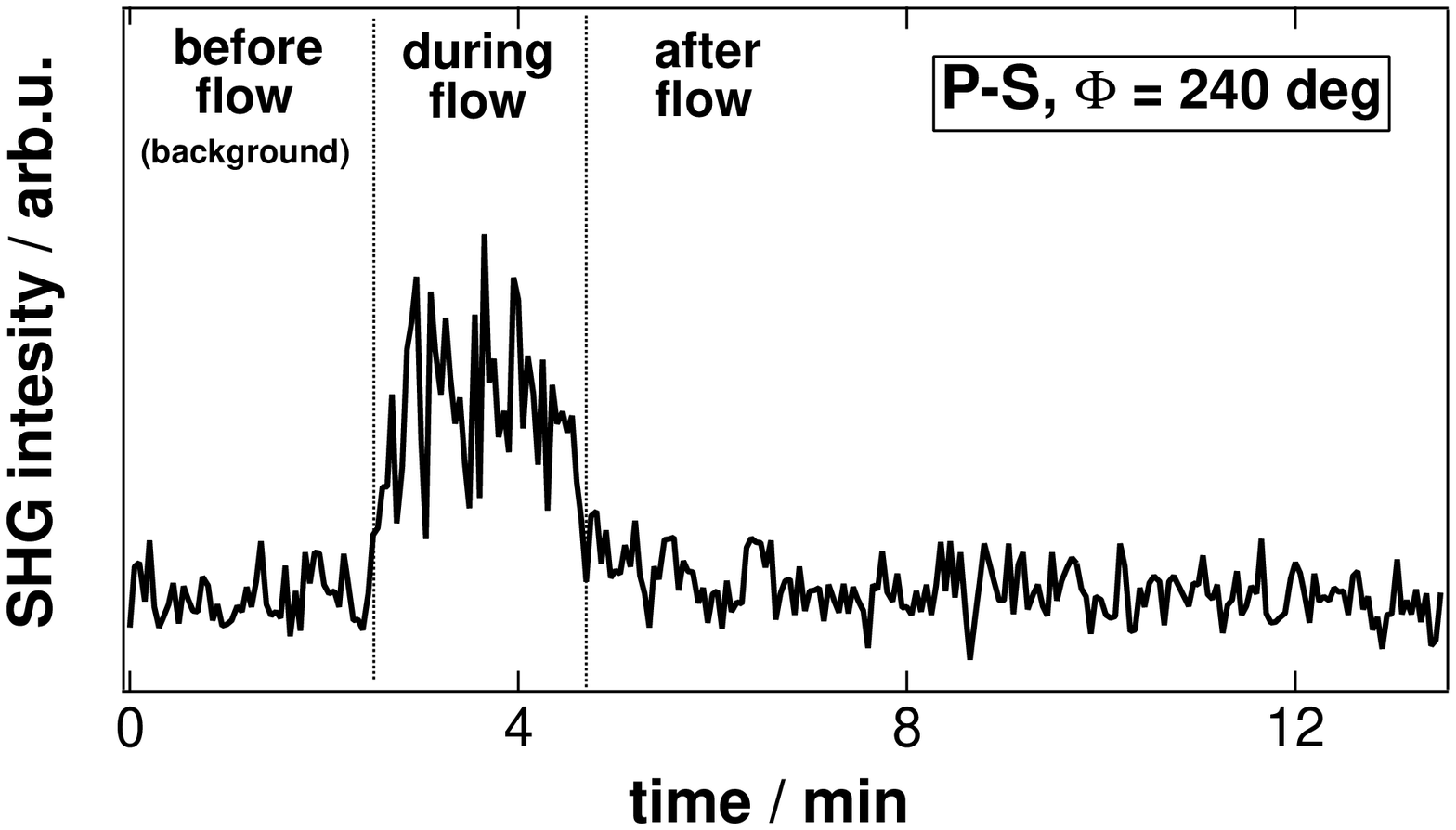, width=0.7\textwidth}
\caption{Example of datafile.
SHG is monitored before, during, and after filling.
Polarization combination and azimuthal sample orientation with respect 
to the flow direction, $\Phi$, are specified.
\label{example_data}
}
\end{center}
\end{figure}

The filling and relaxation processes were monitored with a 
Nd:YAG laser ($\lambda_{\omega} = 1064$\,nm, $\lambda_{2\omega} = 
532$\,nm).
Angles and coordinate system are shown in Figure \ref{schemaSHGflow}.
The SH output was detected by a photomultiplier.
A typical output data-file is shown in Figure \ref{example_data}.
The procedure was repeated for all in-out polarization combinations,
and for several angles $\Phi$.
Since once a cell is filled it cannot be used any longer, a different 
cell was used for each measurement. 
The cell thickness were measured previos filling with a 
spectrophotometer ($\pm 0.1\,\mu$m).

\section{Results}
\begin{figure}
\begin{center}
\epsfig{file=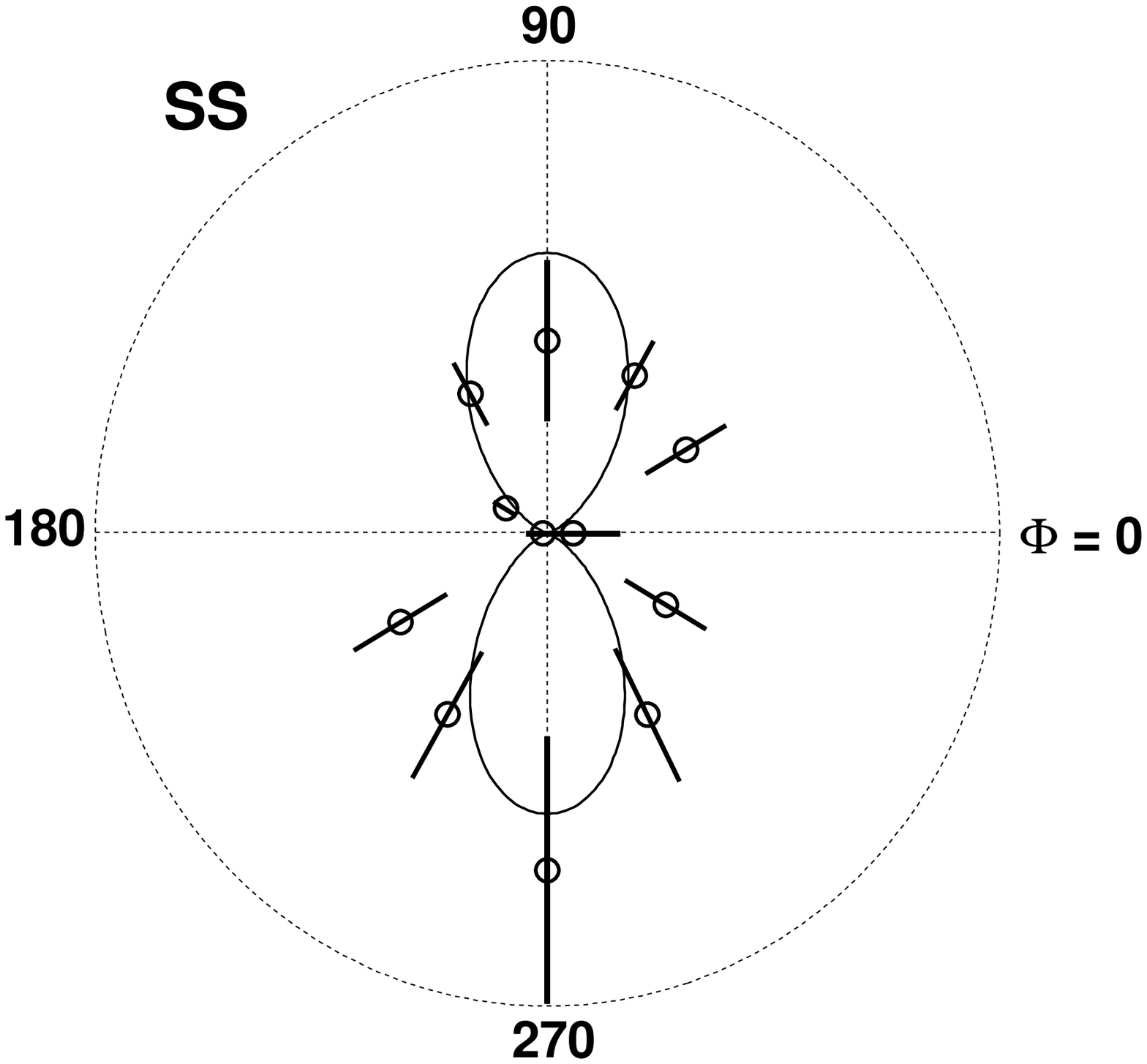, width=0.45\textwidth} 
\hspace{0.04\textwidth}
\epsfig{file=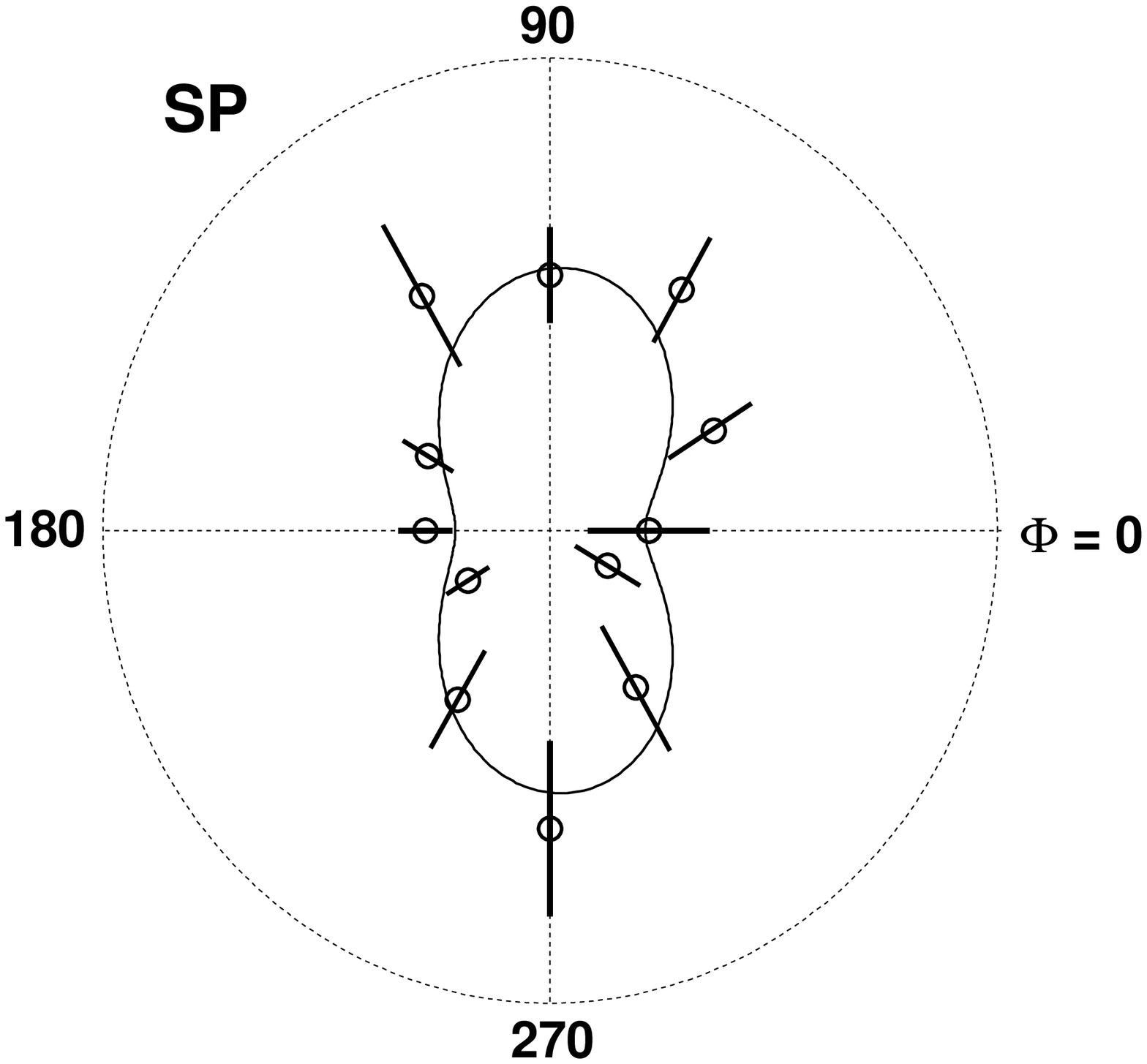, width=0.45\textwidth}\\
\epsfig{file=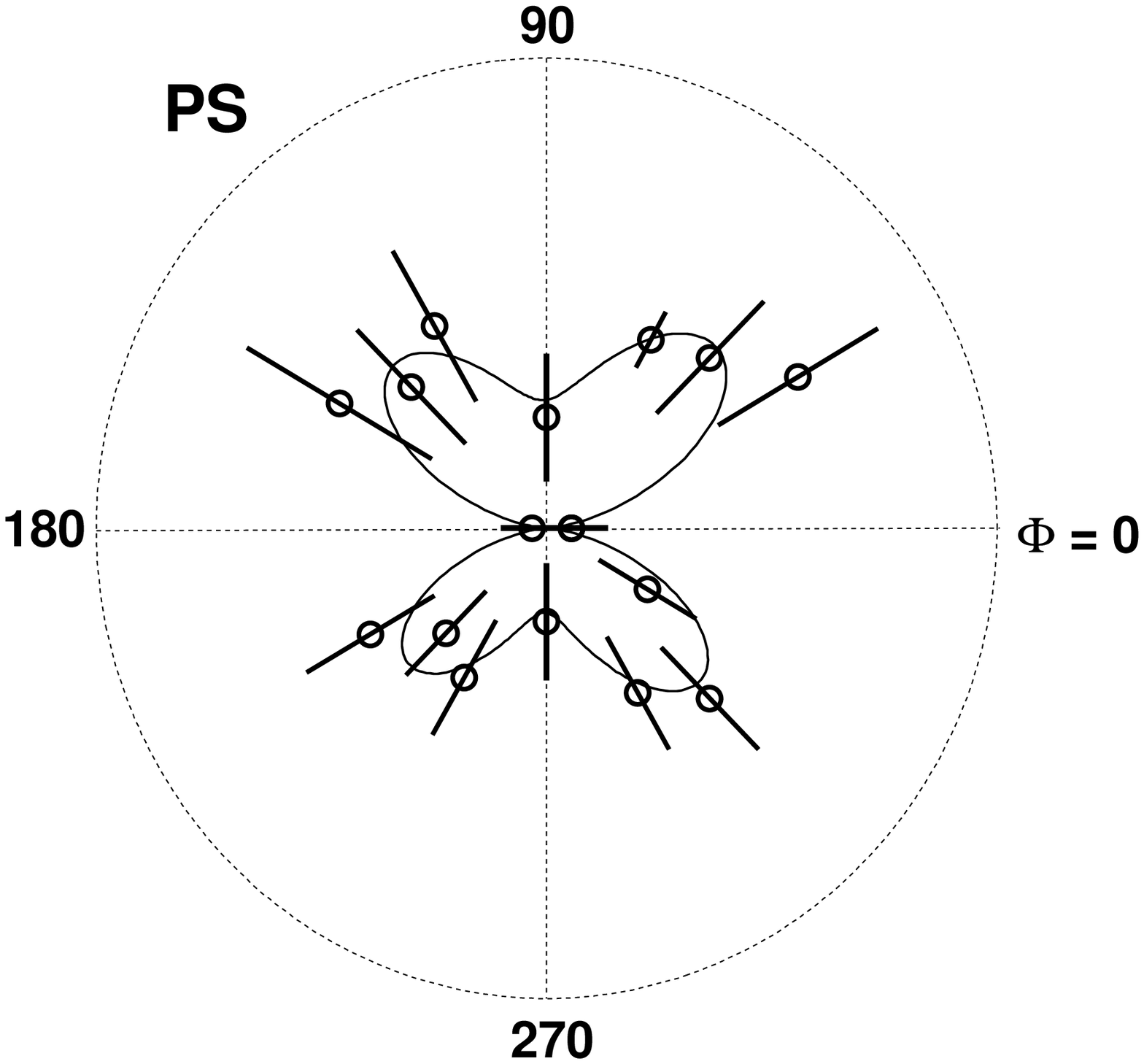, width=0.45\textwidth}
\hspace{0.04\textwidth}
\epsfig{file=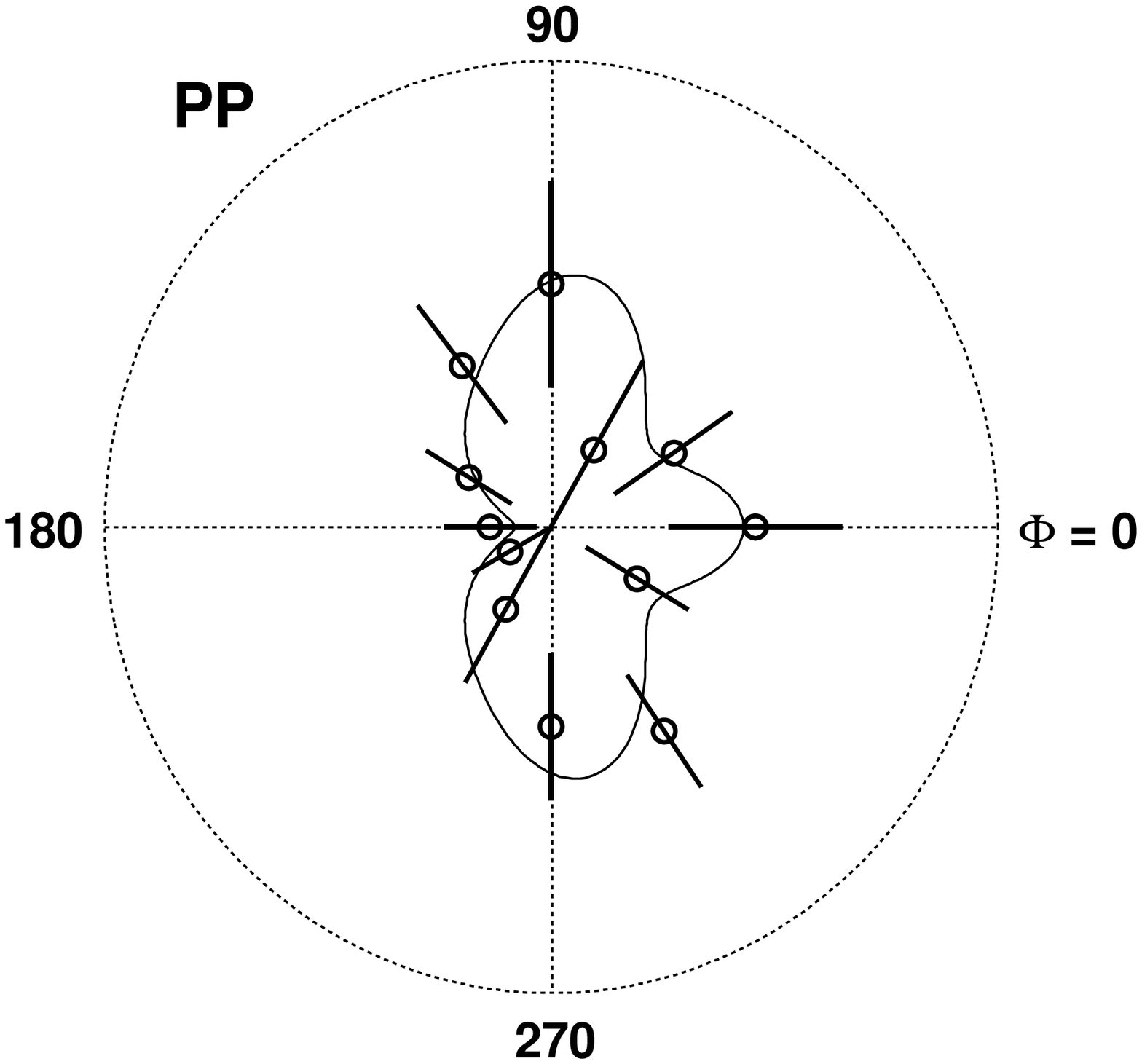, width=0.45\textwidth}\\
\caption{SHG signal (arbitraty units) as a function of the angle $\Phi$ 
(Figure \ref{schemaSHGflow}) during 5CB capillary flow. 
$\Phi = 0$ corresponds to the flow direction.
The open circles are the data with error bars, and the lines are the 
fits following \protect\cite{FelCheShe91}.
The input-output polarizations are shown.
S-polarization is perpendicular to the plane of incidence.
P-polarization is parallel to the incidence plane.
The thickness of the cells used was 10\,$\mu$m.
\label{SHGflow}
}
\end{center}
\end{figure}

Figure \ref{SHGflow} shows the intensity of the SH light generated 
during capillary flow of 5CB at room temparature (nematic phase) 
as a function of the angle $\Phi$ between the flow direction, 
$\mathbf{x}$ 
(Figure \ref{schemaSHGflow}), and the plane of incidence for the four 
different polarization combinations.
All signals show a substancial anisotropy.
The different SH response for $\Phi = 0$ and $\Phi = 180$ in the PP 
plot reflects the preferential alignment of the molecules along the 
filling direction, that is, the LC molecules are tilted preferencially 
along the flow direction.

\begin{figure}
\begin{center}
\epsfig{file=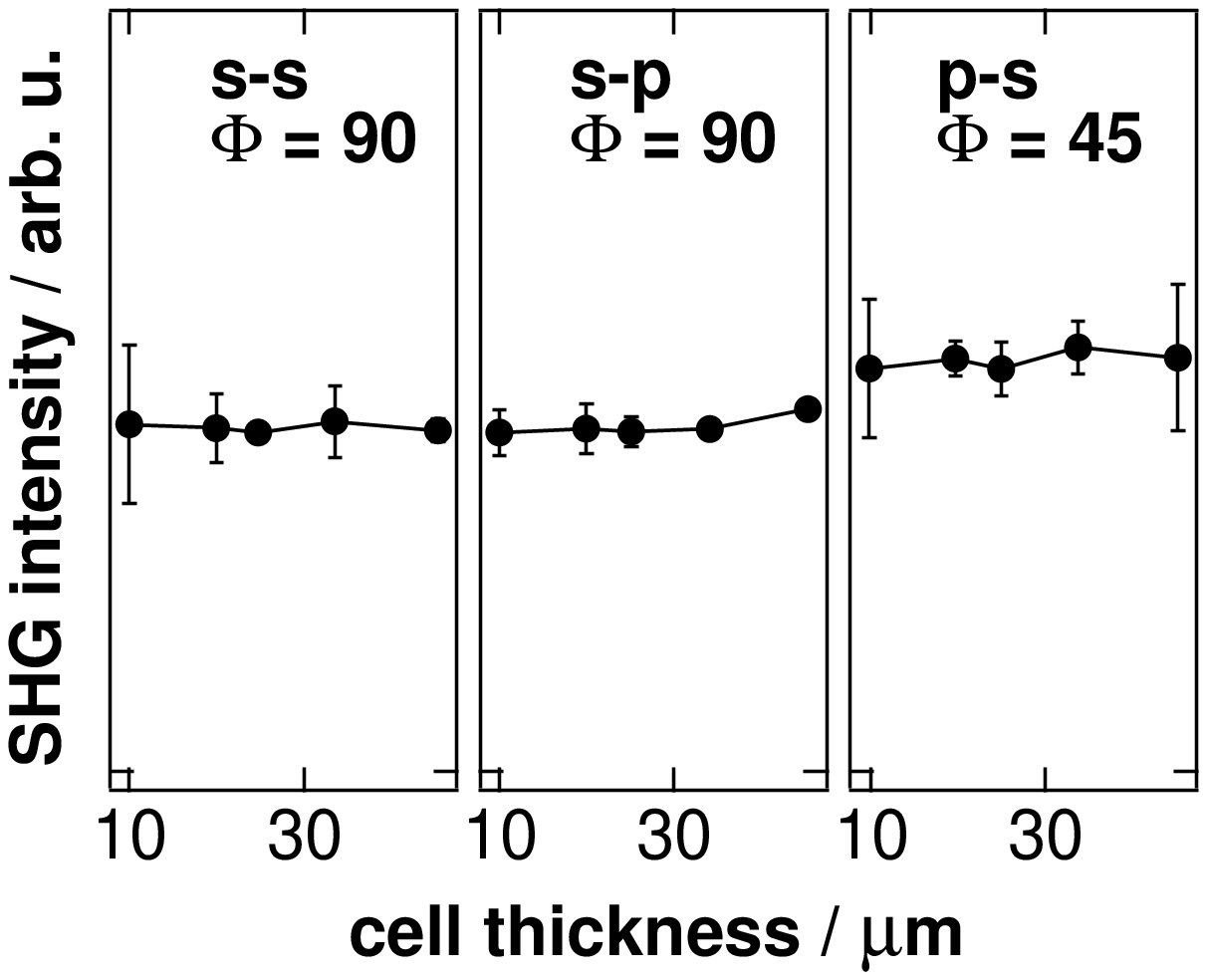, width=0.65\textwidth} 
\caption{SHG intensity as a function of the cell thickness for three 
polarization combination during capillary flow.
The angle between the flow and the incidence plane was chosen in 
order to have the maximum intensity.
The units are the same as in Figure \protect\ref{SHGflow}.
\label{SHG_thick}
}
\end{center}
\end{figure}

The electric quadrupole nonlinearity of the LC might also contribute 
to the SHG\cite{GuyShe88}. 
This contribution is generally many orders of magnitude smaller that 
the dipolar one and proportional to the thickness of the nematic 
layer\cite{JerShe93,HarIbnMohMot98}. 
To ensure that the SH signal of Figure \ref{SHGflow} was only coming
from the NLC surface monolayer the SHG intensity was measured as a function 
of the cell thickness for some polarization combinations
(Figure \ref{SHG_thick}).

\begin{figure}
\begin{center}
\epsfig{file=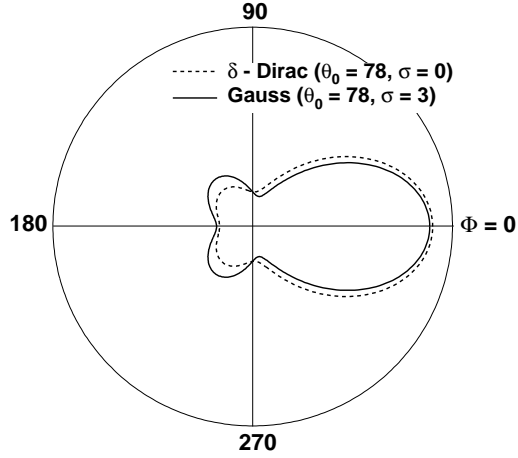, width=0.5\textwidth}
\caption{Azimuthal orientational distribution function 
$g_{\phi,\text{flow}}$ of the first NLC layer during 
capillary flow.
\label{g_flow}
}
\end{center}
\end{figure}

The data in Figure \ref{SHGflow} could be fitted following 
\cite{FelCheShe91} to find the values of the second-order nonlinear 
susceptibilities that yeald the coefficients of the distribution 
functions $f_{\theta}$ and $g_{\phi}$ (Equations \ref{50}-\ref{60} and 
\ref{90}, respectively).
The calculated values of $\theta_{0,\text{flow}}$, $\sigma_{\text{flow}}$, 
and of the $a_{n,\text{flow}}$ coefficients are listed in Table 
\ref{table_flow}.
\begin{table}
\caption{Polar tilt angle and azimuthal orientation distribution 
coefficients of 5CB during flow.
\label{table_flow}
}
\begin{center}
\begin{tabular}{cccccc}
\hline
\hline
Distribution  & $\theta_{0,\text{flow}}$ [deg] & $\sigma_{\text{flow}}$ 
[deg] & $a_{1,\text{flow}}$ & $a_{2,\text{flow}}$ & $a_{3,\text{flow}}$\\
\hline
$\delta(\theta)$ & 78 & 0 & 0.121 & 0.081 & 0.044 \\
Gaussian & 78 & 3 & 0.098 & 0.081 & 0.062 \\
\hline
\hline
\end{tabular}
\end{center}
\end{table}
The azimuthal orientational distribution $g_{\phi,\text{flow}}$ of the 
monolayer during capillary flow was calculated from the $a_{n,\text{flow}}$ 
coefficients listed in Table \ref{table_flow} and is shown in Figure 
\ref{g_flow}.
The distribution is highly anisotropic and shows that flow causes the 
molecules to lie preferentially parallel to the flow direction with a 
pretilt angle $\theta_{0,\text{flow}} = 78$ degrees.
The surface order parameter, as calculated from $a_{2,\text{flow}}$, 
is quite small, $S_{\text{s},\text{flow}} = 0.04$.

\begin{figure}
\begin{center}
\epsfig{file=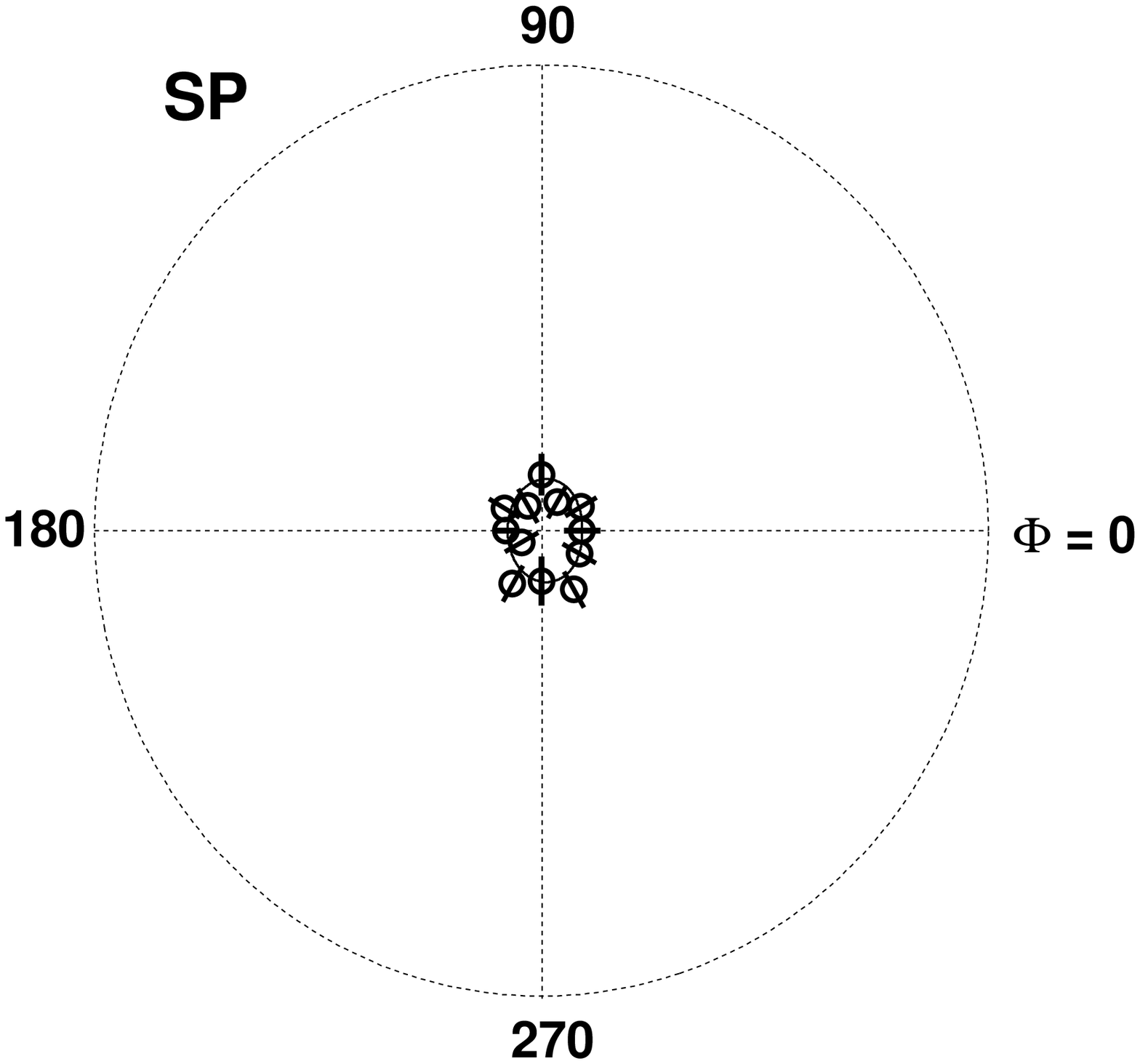, width=0.41\textwidth}
\hspace{0.06\textwidth}
\epsfig{file=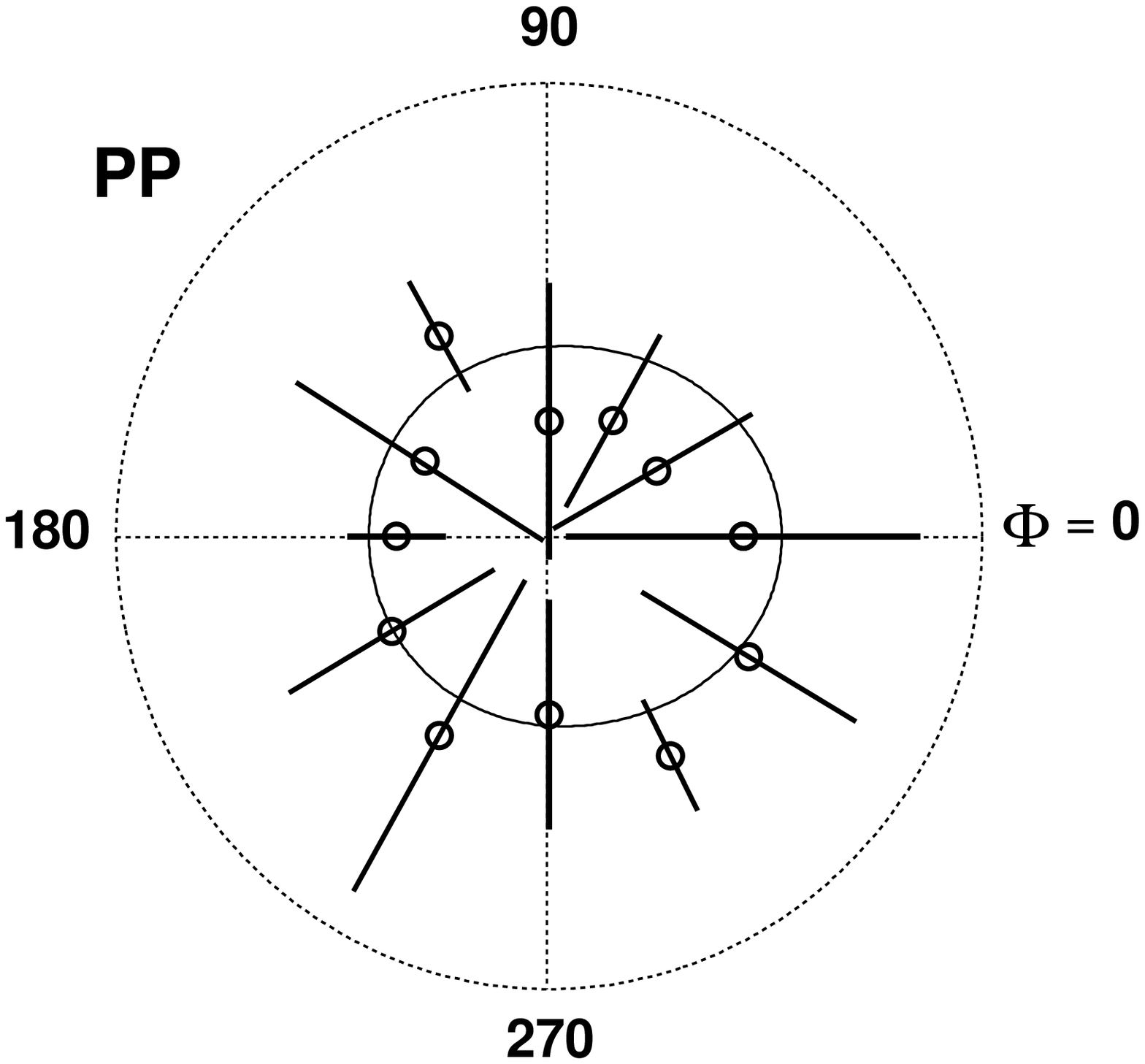, width=0.45\textwidth}
\caption{SHG signal (arbitrary units) vs $\Phi$ angle from a 5CB cell 
after flow. 
Only SP and PP polarization combination generated detectable signal. 
\label{SHGafterflow}
}
\end{center}
\end{figure}

After the capillary flow has ceased the metastable quasi-planar
alignment relaxes to homeotropic and the SHG signal changes 
(as in the Figure \ref{example_data}).
A measurable signal, almost isotropic in $\Phi$, was detected for SP and 
PP polarization combinations (Figure \ref{SHGafterflow}), while no detectable 
signal was generated for SS and PS polarization combinations.
These two facts suggest that, after relaxation, the monolayer is 
isotropically oriented in the $xy$ plane.
S-out conversions are forbidden by symmetry in such a case 
\cite{FelCheShe91}, whereas the allowed conversions should produce
a signal which is azimuthally isotropic.

\begin{figure}
\begin{center}
\epsfig{file=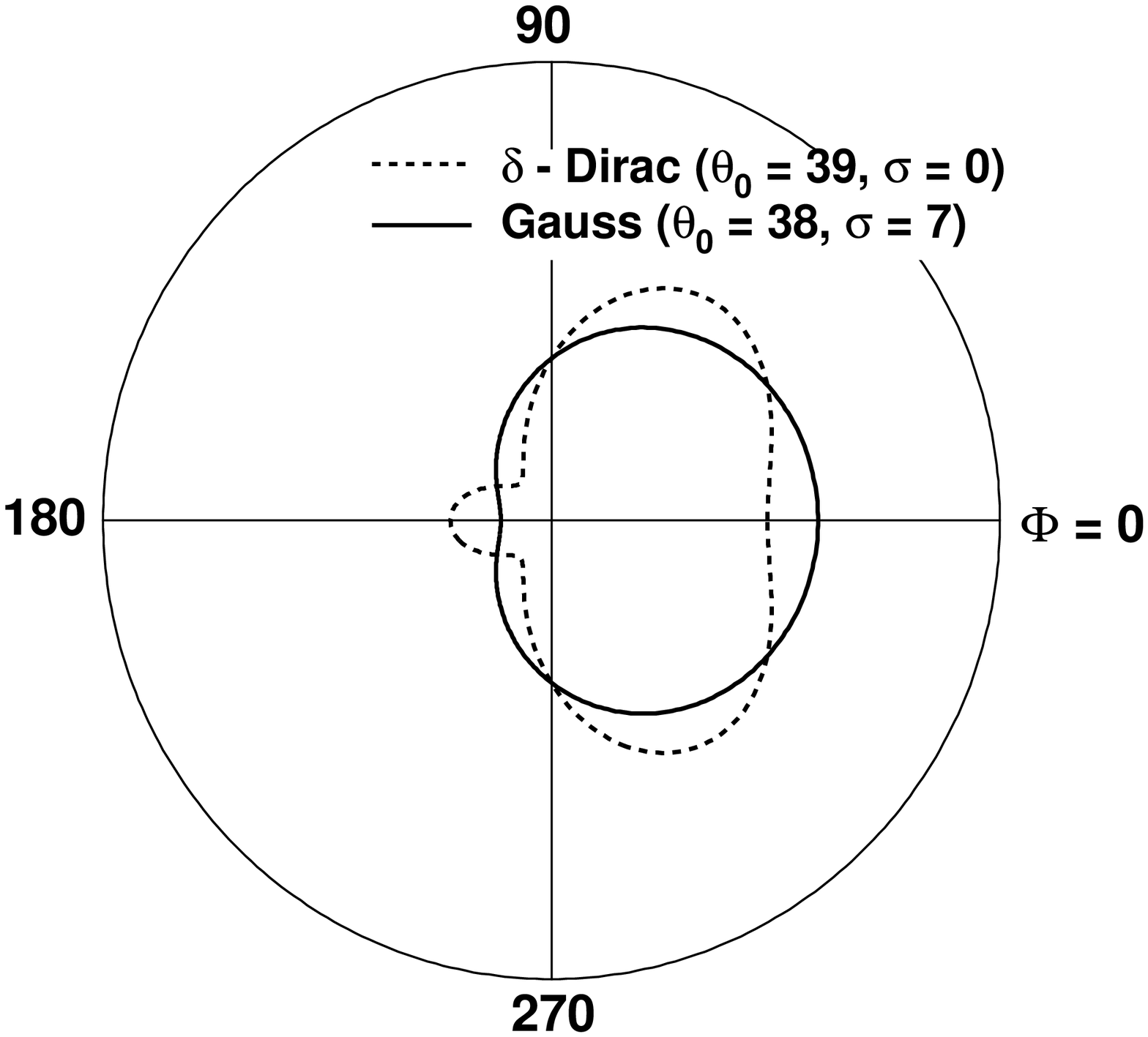, width=0.5\textwidth}
\caption{Azimuthal orientational distribution function 
$g_{\phi,\text{hom}}$ 
of the NLC molecules in the first layer after the flow has ceased and 
the nematic layer has relaxed from the flow-induced alignment to the 
homeotropic one.
\label{g_after}
}
\end{center}
\end{figure}

The values of the tilt angle $\theta_{0,\text{hom}} \pm \sigma_{\text{hom}}$, 
and of the $a_{n,\text{hom}}$ coefficients calculated from the fits in Figure 
\ref{SHGafterflow} are listed in Table \ref{table_afterflow} and the 
azimuthal distribution function $g_{\phi,\text{hom}}$ of the relaxed LC monolayer 
is shown in Figure \ref{g_after}.
\begin{table}
\caption{Polar tilt angle and azimuthal orientation distribution 
coefficients of 5CB after relaxation to the bulk homeotropic alignment.
\label{table_afterflow}
}
\begin{center}
\begin{tabular}{cccccc}
\hline
\hline
Distribution  & $\theta_{0,\text{hom}}$ [deg] & $\sigma_{\text{hom}}$ [deg] 
& $a_{1,\text{hom}}$ & $a_{2,\text{hom}}$ & $a_{3,\text{hom}}$\\
\hline
$\delta(\theta)$ & 39 & 0 & 0.107 & 0 & 0.049 \\
Gaussian & 38 & 7 & 0.111 & 0 & -0.010 \\
\hline
\hline
\end{tabular}
\end{center}
\end{table}
The coefficient $a_{2,\text{hom}}$ of the relaxed monolayer is zero, 
which implies that the in-plane order parameter 
$S_{s,\text{hom}}=0$.
A zero surface order parameter corresponds to an isotropic azimuthal
distribution of the NLC molecules in in the monolayer.
However, the value of $S_{s,\text{hom}}$ is questionable 
because it is within the experimental acciracy of our setup.
We can only conclude that $S_{s,\text{hom}}$ must indeed be 
very small.

The azimuthal distribution function $g_{\phi,\text{hom}}$ shows memory 
of the flow: the NLC molecules are tilted an angle 
$\theta_{0,\text{hom}}=39$ degrees almost isotropically, with a slight 
preference for the flow direction ($\Phi$ = 0). 

The relaxed state was studied by conoscopy and the homeotropic 
alignment was found to be very good.

\section{Discussion and conclusions}
\begin{figure}
\begin{center}
\epsfig{file=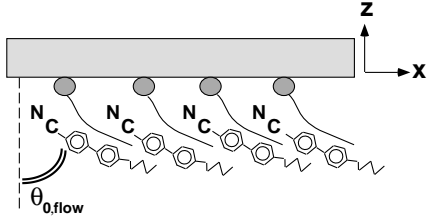, width=0.41\textwidth}
\caption{Sketch of the aligning stearic acid monolayer and of the 
first monolayer of 5CB molecules during capillary flow.
The NLC (LB) molecules are tilted an angle $\theta_{0,\text{flow}}$
in the azimuthal direction of the flow.
\label{tilt}
}
\end{center}
\end{figure}

During capillary flow the first NLC monolayer in contact with the 
aligning LB film results to be oriented in the direction of the flow 
with a quite large pretilt ($\theta_{0,\text{flow}} = 78$ degrees).
If the CN groups of the 5CB molecules are oriented towards the 
surface, then during flow the molecules are tilted as in Figure 
\ref{tilt}.
This suggests that the LB film itself is affected by the flow.
Once the flow ceases the stearic acid alkyl chains relax to their 
equilibrium orientation with a pretilt $\theta_{0,\text{hom}} = 38$ 
degrees almost isotropically distributed in the cell plane, with a 
preference for the flow direction.
The NLC monolayer in contact with the surface follows the fast relaxation 
of the LB film and propagates the homeotropic alignment into the LC 
bulk via the long-range nematic and the elastic interactions.
This interpretation is in agreement with the observed dependence of 
the relaxation speed on the LB monolayer \cite{FazNanKomXX}.

In polarizing microscopy one observes circular domains of 
homeotropic alignment expanding in the quasi-planar structure (see also 
Ref. \cite{FazKomLag98a}).
This regular circular form is indication of the isotropic character 
of the surface relaxation process, which is confirmed by the isotropic 
azimuthal distribution of the NLC molecules after realaxation.
Even though after relaxation the LB film presents a substancial pretilt,
the NLC bulk is homeotropically oriented because the pretilt is 
distributed almost isotropically (conical degeneracy) in the cell plane 
and the overall effect on the LC bulk is that of inducing homeotropic 
alignment.

Some authors refer to this phenomenon as subsurface deformation
\cite{BarEvaPon96, SkaAleBarZum98}. 
In general, the bulk NLC alignment differs from that of the first 
monolayer(s) and a subsurface distortion exists for any tilted anchoring
\cite{RajBarGalOld96}. 
However, the analysis is made assuming that the surface tilt is 
confined to the $xz$ plane and that the anchoring at the surface is 
strong (infinite).
In this case the largest distortion, estimated to be about 1--2 degrees
of polar orientation, is localized over the first molecular layers and 
it has a tail decaying as the inverse distance from the surface which 
makes it to extend over a microscopic length.
However, NLCs in general align weakly on surfactants.
The anchoring strengths are on the order of 10$^{-6}$\,J m$^{-2}$
\cite{BlinovChigrinov}, up to three orders of magnitude smaller than
what is generally considered strong anchoring.
For such weak anchorings one would expect the surface and the bulk 
orientations to differ more substancially than in the case of strong 
anchoring, which is in fact what we observed in this work (the bulk 
and surface tilts differ by 39 degrees).

The LB monolayer preserves a memory of the flow, which is visible 
through the anisotropic azimuthal distribution of the first 
nematic layer, which shows that the surfactant molecules have a slight 
preference to be oriented in the direction of the previous flow.
This flow memory is not directly visible in the equilibrium homeotropic 
alignment, but it has been observed in the case of MBBA 
homeotropically aligned by stearic acid\cite{FazKom99}: 
after electric-field--induced breaking of the
anchoring the homeotropic alignment is not immediately restored, but 
the sample presents a quasi-planar orientation along the direction of 
the previous flow.

In conclusion, we have directly shown that capillary flow influences 
considerably the orientation of the first monolayer of nematic liquid 
crystal in contact with the surfactant LB film, and, indirectly, that 
the flow affects the surfactant layer as well.

\section{Acknowledgments}
Dr. D. Paparo and Dr. L. Marrucci are gratefully 
acknowledged for the very helpfull discussions. 
V.S.U. Fazio acknowledges the financial support of the European TMR 
Programme (contract number ERBFMBICT983023).

%
%
%\section{Articles not (yet) included in the article}
%\cite{ZhuMarJohShe95,JerOBrOucSta93,OBrMosCheFre93,OBrMosCheFre93,
%HarIbnMohMot98,AleBarBarBon95,AleBarPon96,AleBarIonZve96}

%
\bibliography{journal2,flow}

\begin{thebibliography}{10}

\bibitem{GuyHsiShe86}
P.~{Guyot-Sionnest}and H.~{Hsiung} and Y.~R. {Shen}.
\newblock Surface polar ordering in a liquid crystal observed by optical
  second-harmonic generation.
\newblock {\em Phys. Rev. Lett.}, 57(23):2693 -- 2696, 1986.

\bibitem{SkaAleBarZum98}
G.~{Ska\u cej}, A.~L. {Alexe}-{Ionescu}, G.~{Barbero}, and S.~{\u Zumer}.
\newblock Surface-induced nematic order variation: Intrinsic anchoring and
  subsurface director deformations.
\newblock {\em Phys. Rev.~E}, 57(2):1780 -- 1788, 1998.

\bibitem{Jerome91}
B.~{J\'er\^ome}.
\newblock Surface effects and anchoring in liquid crystals.
\newblock {\em Rep. Prog. Phys.}, 54:391 -- 451, 1991.

\bibitem{deGennes}
P.~G. de~{Gennes}.
\newblock {\em The physics of liquid crystals}.
\newblock {Oxford} {University} {Press}, {Oxford}, 1974.

\bibitem{ZhuMarShe94}
X.~{Zhuang}, L.~{Marrucci}, and Y.~R. {Shen}.
\newblock Surface-monolayer-induced bulk alignment of liquid crystals.
\newblock {\em Phys. Rev. Lett.}, 73(11):1513 --1516, 1994.

\bibitem{PrasadWilliams}
P.~N. {Prasad} and D.~J. {Williams}.
\newblock {\em Introduction to nonlinear optical effects in molecules and
  polymers}.
\newblock {John} {Wiley} \& {Sons}, New York, 1991.

\bibitem{Shen}
Y.~R. {Shen}.
\newblock {\em The principles of nonlinear optics}.
\newblock Wiley, New York, 1984.

\bibitem{BarChuKre92}
G.~{Barbero}, A.~N. {Chuvyrov}, and A.~P. {Krekhov}.
\newblock Influence of the flow on the orientation induced by a solid substrate
  on a nematic liquid crystal.
\newblock {\em Int. J. of Mod. Phys. B}, 6(3 \& 4):437 --448, 1992.

\bibitem{FazKomLag98a}
V.~S.~U. {Fazio}, L.~{Komitov}, and S.~T. {Lagerwall}.
\newblock Alignment and alignment dynamics of nematic liquid crystals on
  {Langmuir}-{Blodgett} mono-layers.
\newblock {\em Liq. Cryst.}, 24(3):427 -- 433, 1998.
\newblock http://xxx.uni-augsburg.de/abs/cond-mat/9707043.

\bibitem{FazNanKomXX}
V.~S.~U. {Fazio}, F.~{Nannelli}, L.~{Komitov}, and S.~T. {Lagerwall}.
\newblock Sensitive methods for estimation of the anchoring strength of nematic
  liquid crystals on {Langmuir}-{Blodgett} monolayers of fatty acids.
\newblock {\em Submitted}, 2000.
\newblock http://xxx.uni-augsburg.de/abs/cond-mat/0010484.

\bibitem{FazKomLag98b}
V.~S.~U. {Fazio}, L.~{Komitov}, and S.~T. {Lagerwall}.
\newblock Alignment of nematic liquid crystals on mixed {Langmuir}-{Blodgett}
  mono-layers.
\newblock {\em Thin Solid Films}, 327--329:681 -- 685, 1998.
\newblock http://xxx.uni-augsburg.de/abs/cond-mat/9709180.

\bibitem{FazKom99}
V.~S.~U. {Fazio} and L.~{Komitov}.
\newblock Alignment transition in a nematic liquid crystal due to field-induced
  breaking of anchoring.
\newblock {\em Europhys. Lett.}, 46(1):38 -- 42, 1999.
\newblock http://xxx.uni-augsburg.de/abs/cond-mat/9810293.

\bibitem{note00}
For \lq\lq molecular orientation\rq\rq\, we mean the orientation of the
  cyanobyphenil core of the liquid crystal molecule, and not that of the whole
  molecule; the molecular core is the only part of the molecule whose
  orientation can be probed by second-harmonic generation, because it is the
  only part possessing a a nonlinear hiperpolarizability.

\bibitem{Boyd}
R.~W. {Boyd}.
\newblock {\em Nonlinear optics}.
\newblock {Academic} {Press} {Inc.}, Boston, 1992.

\bibitem{FelCheShe91}
M.~B. {Feller}, W.~{Chen}, and Y.~R. {Shen}.
\newblock Investigation of surface-induced alignment of liquid-crystal
  molecules by optical second-harmonic generation.
\newblock {\em Phys. Rev.~A}, 43(12):6778 -- 6792, 1991.

\bibitem{note01}
As an example, in \cite{JagemalmPontus99} it was found that the polar and
  azimuthal distribution functions of a nematic oriented by obliquely
  evaporated silicon oxide (two-fold degenerate anchoring) are not independent.

\bibitem{KaiNakKan87}
A.~{Kaito}, N.~{Nakayama}, and H.~{Kanetsuna}.
\newblock Infrared dichroism and visible-ultraviolet dichroism studies on
  roller-drawn polypropylene and polyethilene sheets.
\newblock {\em J. Macromol. Sci.--Phys. B}, 26(3):281 -- 306, 1987.

\bibitem{BarAerHolDam92}
M.~{Barmentlo}, N.~A. J.~M. van {Aerle}, R.~W. {Hollering}, and J.~P.~M.
  {Damen}.
\newblock Surface induced liquid-crystla alignment studied by optical
  second-harmonic generation.
\newblock {\em J. Appl. Phys.}, 71(10):4799 -- 4804, 1992.

\bibitem{GuyShe88}
P.~{Guyot-Sionnest} and Y.~R. {Shen}.
\newblock Bulk contribution in surface second-harmonic generation.
\newblock {\em Phys. Rev.~B}, 38(12):7985 -- 7989, 1988.

\bibitem{JerShe93}
B.~{J\'er\^ome} and Y.~R. {Shen}.
\newblock Anchoring of nematic liquid crystals on mica in the presence of
  volatile molecules.
\newblock {\em Phys. Rev.~E}, 48(6):4556 --4574, 1993.

\bibitem{HarIbnMohMot98}
M.~{Harke}, M.~{Ibn}-{Elhai}, H.~{M\"ohwald}, and H.~{Motschmann}.
\newblock Polar ordering of smectic liquid crystals within the interfacial
  region.
\newblock {\em Phys. Rev.~E}, 57(2):1860 -- 1811, 1998.

\bibitem{BarEvaPon96}
G.~{Barbero}, L.~R. {Evangelista}, and S.~{Ponti}.
\newblock Subsurface deformations in nematic liquid crystals.
\newblock {\em Phys. Rev.~E}, 53(1):1265 -- 1268, 1996.

\bibitem{RajBarGalOld96}
M.~{Rajteri}, G.~{Barbero}, P.~{Galatola}, C.~{Oldano}, and S.~{Faetti}.
\newblock Van der waals-induced distortions in nematic liquid crystals close to
  a surface.
\newblock {\em Phys. Rev.~E}, 53(6):6093 -- 6100, 1996.

\bibitem{BlinovChigrinov}
L.~M. {Blinov} and V.~G. Chigrinov.
\newblock {\em Electrooptic effects in liquid crystal materials}.
\newblock Springer Study, New York, 1995.

\bibitem{JagemalmPontus99}
Pontus {J\"agemalm}.
\newblock {\em On the optics and surface physics of liquid crystals}.
\newblock PhD thesis, {G\"oteborg} {University} and {Chalmers} {University} of
  {Technology}, {G\"oteborg}, {Sweden}, 1999.

\end{thebibliography}
\end{document}